\documentclass[ap]{emulateapj}
\usepackage{mathtools}
\usepackage{amsmath, bm}
\usepackage{hyperref}
\usepackage{physics}

\hypersetup{
    colorlinks=true,
    linkcolor=blue,
    anchorcolor = red,
    citecolor = blue,
    filecolor = red,
    pagecolor = red,
    urlcolor = blue
}

\begin{document}
\title{Large-scale dynamics of winds originated from black hole accretion flows: (I) Hydrodynamics}
 
\author
{Can Cui$^{1,2}$, Feng Yuan$^{1,2}$, and Bo Li$^{3}$}
\affil{$^{1}$Shanghai Astronomical Observatory, Chinese Academy of Sciences, Shanghai 200030, China }
\affil{$^{2}$University of Chinese Academy of Sciences, 19A Yuquan Road, Beijing 100049, China}
\affil{$^{3}$Institute of Space Sciences, Shandong University, Weihai 264209, China}
\begin{abstract}

Winds from black hole accretion flows are ubiquitous. Previous works mainly focus on the launching of wind in the  accretion flow scale. It still remains unclear how far the winds can propagate outward and what is their large-scale dynamics.  As the first paper of this series, we study the large-scale dynamics of thermal wind beyond accretion scales via analytical and numerical methods. Boundary conditions, which are crucial to our problem, are analyzed and presented based on the small-scale simulations combined with observations of winds. Both black hole and galaxy potential are taken into account.  For winds originated from hot accretion flows, we find that the wind can reach to large scales. The radial profiles of velocity, density, and temperature can be approximated by $v_r\approx v_{r0}, \rho\approx \rho_{0}(r/r_0)^{-2}$, and $T\approx T_0 (r/r_0)^{-2(\gamma-1)}$, where $v_{r0}, \rho_0, T_0$ are the velocity, density, and temperature of winds at the boundary $r_0(\equiv 10^3 r_g)$, $\gamma$ is the polytropic index. During the outward propagation, the enthalpy and the rotational energy compensate the increase of gravitational potential. For thin disks, we find that because the Bernoulli parameter is smaller, winds cannot propagate as far as the hot winds, but stop at a certain radius where the Bernoulli parameter is equal to the potential energy. Before the winds stop, the profiles of dynamical quantities can also be approximated by the above relations. In this case the rotational energy alone compensates the increase of the potential energy. 

\end{abstract}

\section{Introduction} \label{sec:intro}

Observational evidence of winds from accretion disks has been accumulating for both cold and hot accretion flows over the past two decades. Cold accretion disks produce high velocity winds, which have been widely observed in luminous AGNs (e.g. \citealp{crenshaw_etal03,tombesi_etal10,liu_etal13,tombesi_etal14,kp15}) and black hole X-ray binaries (e.g. \citealp{nh12,homan_etal16,db16}). Detections of winds from hot accretion flows are more challenging since they are usually fully ionized. Nevertheless, wind observations from low-luminosity sources have gradually built up in recent years through various approaches. These detections include low-luminosity AGNs and radio galaxies \citep{tombesi_etal10,tombesi_etal14,ck12,cheung_etal16}, the supermassive black hole in the Galactic center (Sgr A*; \citealp{wang_etal13,ma2019}), and  black hole X-ray binaries in hard state \citep{homan_etal16,munoz2019}.

Theoretically, the launching of winds has been extensively studied. For thin disks, three driving mechanisms have been proposed (\citealp{Proga07}), namely thermal mechanism (e.g. \citealp{Begelman1983}), in either the context of circumstellar disks (e.g. \citealp{font_etal04}) or black hole accretion disks (e.g. \citealp{luketic_etal10,wp12,ca16}), radiation (line force; e.g. \citealp{Murray1995,Proga2000,Proga2004,Nomura2017}), and magnetic field (e.g. \citealp{Blandford1982,Lovelace1994,Romanova1997,Cao2014,bai_etal16}). It is likely that all three mechanisms play a role, though most existing works focus only on one mechanism given its technical difficulty. Two examples of exception are \citet{proga03} and \citet{waters18}, in which both the thermal and magnetic mechanisms are taken into account.

Despite the relative rarity of observational evidence for winds launched from hot accretion flows, the theoretical understanding is more advanced than the case of thin disks. This is attributed partly to that radiation is dynamically unimportant in hot accretion flows and that it is technically easier to simulate geometrically thick flows. It has been long suspected that strong winds should exist in hot accretion flows \citep{narayan1994,blandford1999,stone99,stone01}. This speculation was later confirmed by detailed numerical simulations \citep{yuan_etal12a,yuan_etal12b,narayan_etal12,li_etal13} and analytical studies \citep{bu_etal16a,bu_etal16b,bu18}. 
\citet{yuan_etal15} (hereafter Y15) study the properties of wind originated from hot accretion flows based on three dimensional (3D) general relativistic (GR) magnetohydrodynamical (MHD) simulations, such as velocity and mass flux. They analyze data with a virtual particle trajectory approach which can effectively discriminate real wind from turbulent motions. This approach ensures the validity of wind properties obtained.

Almost all the above-mentioned works focus on the accretion flow scale. On the much larger scale, winds are now widely invoked to play a critical role for the interaction between the active galactic nuclei (AGN) and the host galaxy, resulting in the co-evolution between the two (e.g. \citealp{ciotti_etal10,ciotti_etal17,ostriker_etal10,choi12,weinberger_etal17,eisenreich_etal17,yuan2018,yoon2018,yoon2019}). For example, recently \citet{yuan2018} comprehensively include feedback by wind and radiation  from AGNs in both cold and hot feedback modes and find that wind plays a dominant role in both modes, though radiative feedback cannot be neglected.

Since the scales of wind launching and  feedback differ by orders of magnitude, important questions then emerge: can wind produced in the accretion disk scale escape the gravitational potential of the central black hole and how far can they propagate outward in the combined potential of the black hole and galaxy? What is the detailed dynamics of winds when they propagate outward, i.e., how do the velocity and density of winds change with radius? 

For winds from hot accretion flows, answering these questions is somewhat pressing, because recent cosmological simulations find that to overcome some serious problems in galaxy formation, e.g., reducing star formation efficiency in the most massive halos, winds launched from hot accretion flows must be invoked to interact with the interstellar medium in the galaxy scale (e.g., \citealp{weinberger_etal17}). However, since the comprehensive study to the wind launching in the accretion flow scale was performed only recently (i.e., Y15), the large-scale dynamics of winds has not been investigated yet. This is the primary goal of our present work. 

The crucial factor of studying the large-scale dynamics of winds is to employ correct boundary conditions, because the hydrodynamical equations controlling the wind dynamics is a set of differential equations.   Many works have studied winds from thin disks, either around young stars (e.g., \citealp{font_etal04,bai_etal16}) or black holes (e.g., \citealp{Lovelace1994,Romanova1997,Proga2000,proga03,Proga2004,luketic_etal10,wp12,Cao2014,
ca16,Nomura2017,waters18}), and some works have already extended to very large radii. However, since  these works fail to take into account all the above-mentioned three driving mechanisms, they could not supply realistic boundary conditions. Therefore we need to revisit the large-scale dynamics with realistic boundary conditions.

 In addition to the above-mentioned issues, another important factor is that we should take into account the gravitational potential of the host galaxy because we are interested in the dynamics of wind well beyond the accretion flow scale. The effect of the galaxy potential is hard to estimate without doing detailed calculations. Most previous works do not include this ingredient.

In the present paper, we aim at systematically studying the dynamics of winds  launched from both hot accretion flows and cold thin disks. Realistic boundary conditions will be analyzed and adopted and the galaxy potential will be included. The inner boundary is set at $\sim 10^3 r_g$. Radiation force is neglected. This assumption is reasonable for wind from hot accretion flows since the radiation of accretion flow is weak and the gas is fully ionized. But for wind from a thin disk, radiation force is very likely to play an important role. In addition to simplifying calculation, we hope our ``thermal assumption'' can provide a ``lower limit'' to the dynamics of wind.   It is unclear to the role of magnetic field because the observational constraints on the magnitude and configuration are poor. In this paper we  focus on the case without magnetic field; magnetic field will be taken into account in our next work.  

The paper is structured as follows. In \S2, we present analytical solutions of winds from hot accretion flows and thin disks. We will start with an analogy with the solar wind to illustrate why should we only look for transonic or supersonic solutions (\S2.1). After a brief description of the model setup (\S2.2), we present the gravitational potential of the galaxy (\S2.3) and the equations we use (\S2.4).  In \S2.5 we present  a detailed discussion of how do we adopt proper boundary conditions in both cases of cold and hot accretion flows. The results are described in detail in \S2.6. In \S3, we present 1D and 2D hydrodynamical numerical simulations and compare with the analytic solutions. We summarize and discuss our results in \S4.

\section{Analytical Solutions}

\subsection{Analogy with the solar wind}

To study the large-scale dynamics of wind from accretion disks, solar wind can be a valuable reference. A canonical model describing the dynamics of solar wind is the Parker model \citep{p58,p60}, in which distinct types of solutions are found. Requesting the wind velocity to be subsonic in the near-Sun region, only two solutions are allowed, namely, the transonic solution and the subsonic solution. A transonic solution represents a transition from a subsonic to a supersonic flow by passing through the critical (sonic) point, and the radial velocity tends to increase monotonically with radius. A non-monotonical dependence of the radial speed on heliocentric distance occurs when multiple critical points exist as a result of, say, the super-radial expansion of the flow tube (e.g., \citealp{yp77,cod90,lxc11}). Regardless, a transonic solution is always possible whereby the solution chooses the proper sonic point from the available critical points. The subsonic solution, also called the breeze solution in solar wind problems, has its velocity decreasing at large distances and never passes across the critical point.

The subsonic solution has been discarded conventionally for solar winds mainly because: 1) at large radii the solution yields a finite pressure and density, which cannot be matched to the interstellar gas properties ; 2) in situ measurements of the near-Earth solar wind show that the wind speed far exceeds the local sound speed (e.g., \citealp{abbo_etal16}); 3) the subsonic solution is proved to be unstable to low-frequency acoustic perturbations \citep{velli01}. Consequently, we restrict ourselves to transonic solutions which are physical for disk wind problem, though the subsonic solutions are presented for comparison.

\subsection{Model setup}

We solve for steady-state one dimensional (1D)  hydrodynamical equations in spherical polar coordinates $(r, \theta, \phi)$ following  \citet{abramowicz1981}. The inner boundary is set to be $r_0 = 10^3 r_g$ in order to obtain reliable boundary conditions from small-scale simulations of Y15. Here the gravitational radius is defined as $r_g\equiv GM/c^2$ and $M$ is the black hole mass ($r_g\sim10^{-2}$ pc for  $M=10^8M_{\odot}$). We prescribe wind to propagate along a direction with constant angle $\theta=45^{\circ}$ from the equatorial plane, despite that the specific angular momentum is nonzero in the model. This simplification is well justified in 2D simulations in this paper (\S\ref{sec:2d}). We have experienced with different $\theta$ angles and found it did not alter the results qualitatively.

\subsection{Gravitational potential}
In this work we deal with winds extending to galaxy scales so the gravitational potential of both the central black hole and the host galaxy should be taken into account:
\begin{equation}
\Phi = \Phi_\textrm{BH} + \Phi_\textrm{galaxy}.
\end{equation}
We adopt the usual  pseudo-Newtonian gravitational potential to describe the black hole potential \citep{paczynsky1980}, 
\begin{equation}
\Phi_\textrm{BH}  = -\frac{GM}{r-r_{s}},
\end{equation}
where $r_s\equiv 2GM_{\rm BH}/c^2 = 2r_g $ is the Schwarzchild radius.

Observations show that the stellar and the dark matter components of galaxies are distributed so that their total mass profile is well described by a density distribution $\rho \propto r^{-2}$ over a large radial range (e.g., \citealp{treu2002,treu2004,rusin2005,gavazzi2007,czoske2008,dye2008}). This leads to a simple assumption that the galaxy potential is treated as a flat rotation curve with constant velocity dispersion parameter $\sigma$. Then, the difference of the potential of the host galaxy between $r$ and $r_0$ is written by
\begin{align}
\Delta \Phi_\textrm{galaxy} =\int_{r_0}^{r} \frac{\sigma^2}{r} =\sigma^2 \ln r+C,
\end{align}
where $r = \sqrt{R^2 +z^2}$ denotes the spherical radius and $C$ is a constant. We adopt a velocity dispersion of 200 $\mathrm{km\; s^{-1}}$  which is common to elliptical galaxies (\citealt*{kormendy2013}).

\subsection{Equations}\label{sec:eq}
The conservation of mass and energy are given by \citep{abramowicz1981}
\begin{align}
&\rho v_rr^2 = \dot M,  \label{eq:cont} \\
\frac{1}{2} (v_r^2+v_\phi^2 )+&\frac{c_{s}^2}{\gamma-1}-\frac{GM}{r-r_\textrm{s}} + \sigma ^2\ln r = E.  \label{eq:Be}
\end{align}
where $\dot M$ denotes the mass flux of wind, and Equation \eqref{eq:Be} is the Bernoulli intergral with constant Bernoulli parameter $E$. The rotational velocity $v_\phi$ is related to the specific angular momentum by $l = v_\phi R$. The sound speed is defined as $c_{s}^2 \equiv \partial P/\partial \rho=\gamma K \rho^{\gamma-1}$ where $\gamma$ and $K$ are the polytropic index and polytropic constant, respectively. We adopt $\gamma=4/3$ throughout the analytical section and defer our discussion on the adoption of the value and its physical interpretation in \S\ref{sec:gamma}. Substituting $\rho$ by $c_s$ leads us to rewrite Equation (\ref{eq:cont}) as
\begin{equation} \label{eq:mass1}
c_s^\frac{2}{\gamma-1}v_rr^2 = \dot M'.
\end{equation}
Note that here $\dot M'$ differs from $\dot M$ by a constant coefficient. We drop the superscript prime for the rest of the derivations. After a differentiation of Equations (\ref{eq:Be}) and (\ref{eq:mass1}), one arrives at
\begin{multline}
r^3(v_r-c_s)\bigg(\dv{v_r }{r}-\frac{2}{\gamma-1}\dv{c_s}{r}\bigg) = \\
\frac{l^2}{\sin^2\theta} - l^{\prime2} -\sigma^2r^2 +2r^2v_rc_s,
\label{eq:diff}
\end{multline}
where $l^\prime = GMr^{3/2}/(r-r_s)$. At the sonic point, the wind velocity equals the sound speed so that Equation (\ref{eq:diff}) can be reduced to
\begin{align}\label{eq:cs2}
c_s^2 = v_r^2 = &\frac{1}{2r_\textrm{c}^2}\left (-\frac{l^2}{\sin^2\theta} + l^{\prime2}_{r_\textrm{c}} +\sigma^2r_\textrm{c}^2 \right),
\end{align}
where $r_\textrm{c}$ is the locus of the sonic point. Inserting Equation (\ref{eq:cs2}) into Equations (\ref{eq:Be}), one can solve for $r_\textrm{c}$ as a function of $(E,l^2)$, or more easily we can solve for the angular momentum
 \begin{align}\label{eq:l2E}
l^2(E,r_\textrm{c})= & \frac{2R_c^2}{\frac{1}{2}-n} \cdot \nonumber \\
& \left( E-\frac{(\frac{1}{2}+n) (l^{\prime 2} + \sigma^2r_c^2) }{2r_c^2} + \frac{GM}{r_c-r_s} - \sigma^2\ln r_c \right).
\end{align}
Similarly, inserting Equation (\ref{eq:cs2}) into Equations (\ref{eq:mass1}) yields the angular momentum as a function of $(\dot M,r_\textrm{c})$
 \begin{align}\label{eq:l2Mdot}
l^2(\dot M,r_\textrm{c})= ( l^{\prime 2}_{r_\textrm{c}} + \sigma^2 r^2_c -  \dot M^\frac{1}{2n+1} 2r_c^\frac{4n}{2n+1} )\sin^2\theta.
\end{align}

Inspecting Equations (\ref{eq:l2E}) and (\ref{eq:l2Mdot}), one finds that four unknowns $l,E,\dot M,r_c$ are present in two equations, requiring us to specify two of these variables and to solve for the other two. We estimate $l$ and $E$ based upon the  boundary conditions (\S \ref{sec:anaybc}). Once constants $l$ and $E$ are specified, the loci of the sonic points $r_\textrm{c}$ can be computed via Equation (\ref{eq:l2E}) and then mass fluxes $\dot{M}$ via Equation (\ref{eq:l2Mdot}), hence the velocity and temperature (or sound speed) profiles of the transonic solution via solving Equations (\ref{eq:Be}) and (\ref{eq:mass1}).

Our differential equations have three variables so usually we should supply three boundary conditions to solve the equations, namely  radial velocity ($v_{r0}$), rotational velocity ($v_{\phi0}$), and temperature ($T_0$). They in turn determine the values of $E$, $l$, and $\dot{M}$. However, for a set of arbitrarily given boundary conditions the solution in general does not pass through the sonic point, i.e., it is not a transonic solution.  The sonic condition provides an additional constraint which requires a specific combination of the three quantities at the boundary. In other words,  fine-tuning is required to obtain a transonic solution, i.e., only two of them are free to choose for obtaining a transonic solution.

\subsection{Boundary Conditions}\label{sec:anaybc}

In this section, we discuss boundary conditions employed for winds from hot accretion flows and thin disks respectively, i.e.,  the values of $v_{r0},v_{\phi0}$, and $T_0$ at $r_0=10^3 r_g$.

\subsubsection{Hot accretion flows}
\label{hotboundary}

{
\begin{deluxetable*}{lllllllllll}
\tablecaption{Parameters of analytic solutions \label{tab:anaypara}}
\tablehead{
\colhead{} &
\colhead{$\mathrm{Potential}$} &
\colhead{$E$} &
\colhead{$r_c$} &
\colhead{$r_0$ } &
\colhead{$T_0$} &
\colhead{$v_{r0}$}&
\colhead{$v_{\phi 0}$}&
\colhead{Mach$_0$}&
\colhead{Branch}\\
\colhead{} &
\colhead{} &
\colhead{} &
\colhead{$[r_g]$} &
\colhead{$[r_g]$} &
\colhead{[$\mathrm{K}$]} &
\colhead{[$v_{k0}$]}&
\colhead{[$v_{k0}]$}
}
\startdata
\multicolumn{2}{l}{Hot Accretion Flow}\\
\hline
hot\_bh\_s  & bh& $ 1.76\times10^{-4}$  & &$10^3$ &$1.32\times10^9$ &$0.21$  & 0.5  &       0.44 &sub \\
hot\_bh\_t  & bh &$ 1.66\times10^{-4}$  &$3.83\times10^{3}$ &$10^3$ &$1.30\times10^9$ &$0.22$  & 0.5  &    0.46 &trans  \\
hot\_bhg\_s & bh+g & $1.79\times10^{-4}$  && $10^3$& $1.32\times10^9$ &    $0.21$     &  0.5  &  0.44  &  sub\\
hot\_bhg\_ta  & bh+g&  $1.69\times10^{-4}$  &$3.86\times10^{3}$& $10^3$& $1.30\times10^9$ &    $0.22$     &  0.5  & 0.46  &trans \\
hot\_bhg\_tb   & bh+g& $8.00\times10^{-6}$  & $5.65\times10^{5}$  & $10^3$& $1.12\times10^9$ &$5.8\times10^{-4}$  &0.5  &  $1.33\times10^{-3}$  & trans \\
hot\_bhg\_sup  & bh+g& $1.31\times10^{-2}$  && $500$ &   $10^{10}$ & 3.0   &0.7  &  2.29  & sup \\
\\
\multicolumn{11}{l}{Thin Disk}\\
\hline
cold\_fid & bh+g & $ -1.09\times10^{-4}$  &&$10^3$ &$10^5$ &1.0  & 0.5  &    241 & sup \\
cold\_vr0.2& bh+g& $ -7.92\times10^{-4}$  &&$10^3$ &$10^5$ &0.2  & 0.5  &  48.2   & sup \\
cold\_vr2    & bh+g& $7.54\times10^{-3}$  && $10^3$& $10^5$ &2.0  &  0.5  &  482  &  sup \\
cold\_vphi0.3 & bh+g & $ -2.2\times10^{-4}$  &&$10^3$ &$10^5$ &1.0  & 0.3  &    241 & sup \\
cold\_vphi0.1 & bh+g & $ -2.8\times10^{-4}$  &&$10^3$ &$10^5$ &1.0  & 0.1  &    241 & sup \\
cold\_vphi0.01 & bh+g & $ -2.9\times10^{-4}$  &&$10^3$ &$10^5$ &1.0  & 0.05  &    241 & sup \\
cold\_T4 & bh+g&  $-1.10\times10^{-4}$  && $10^3$& $10^4$ &1.0     &  0.5  & 762  & sup\\
cold\_T6 & bh+g&$-1.09\times10^{-4}$  && $10^3$& $10^6$ &1.0     &  0.5  & 76.2 & sup \\
cold\_T4vr2 & bh+g&$2.02\times10^{-3}$  && $10^3$& $10^4$ &2.0     &  0.5  & 76.2 & sup 
\enddata
\end{deluxetable*}
}

The dynamics of hot accretion flows around black holes are well studied (see \citealp{yuan2014} for a review). Wind properties from hot accretion flows are investigated in Y15 based on 3D GRMHD simulations. It is found that winds can be produced from $r \sim 30 r_g$ up to the outer boundary of the accretion flow, implying that winds at $r_0=10^{3}r_g$ are a combination of those originated from $r\leq10^3 r_0$. Furthermore, winds originated from various radii have different velocities, and they almost keep constant when it propagates outward, implying that the velocity at $r_0$ are diverse. The mass flux is proportional to the radius as $\dot M_\mathrm{wind} \propto r^s$, with $s\approx 1$, indicating that wind properties are dominated by the locally generated ones rather than components originated from smaller radii. The mass flux-weighted poloidal velocity is described by $v_{p} \approx0.2 v_k(r)$, with $v_k(r)$ the Keplerian velocity at $r$.

\begin{figure*}[ht!]
\epsscale{1.05}
\plotone{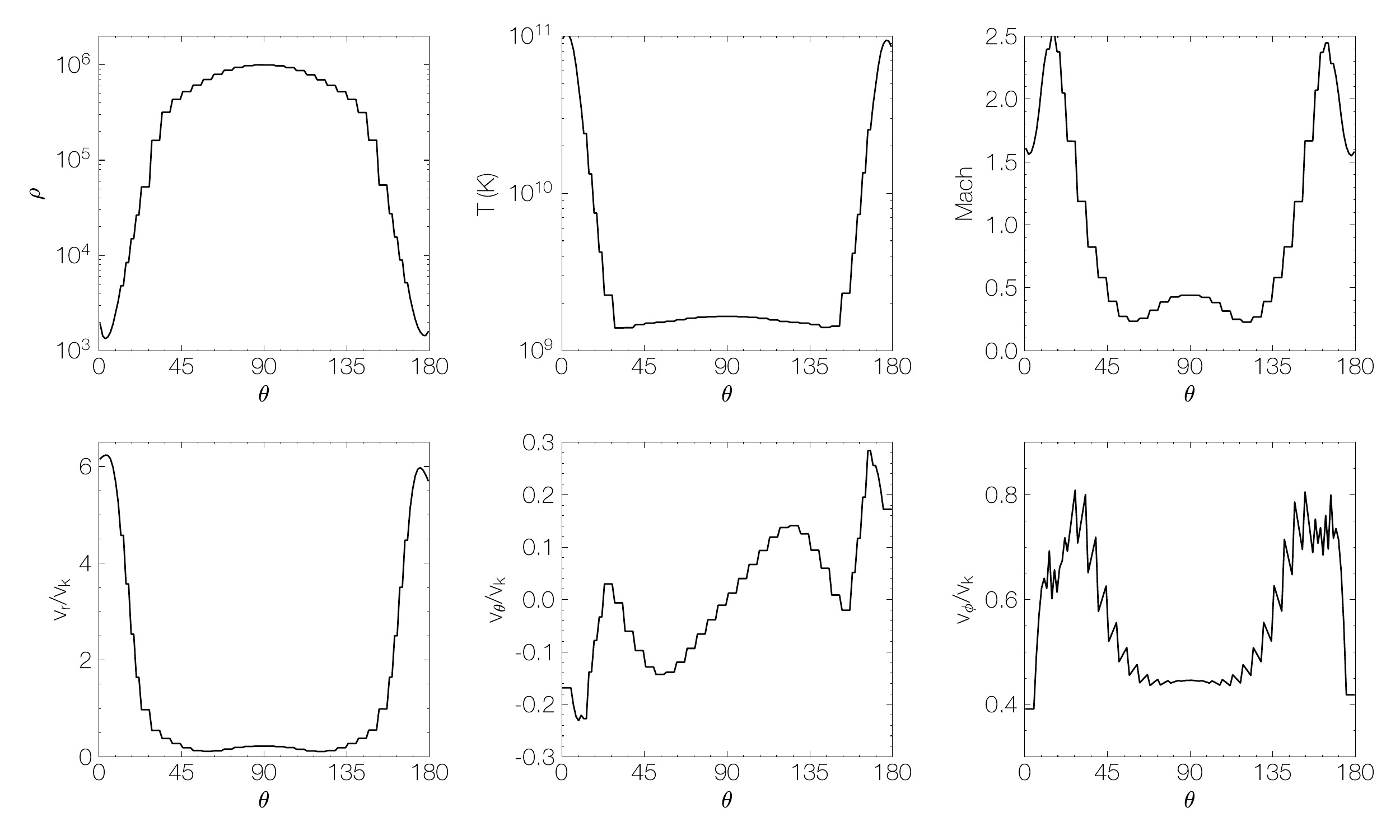}
\caption{Physical quantities of winds as a function of $\theta$ at $r = 10^3r_g$ from the 3D GRMHD  simulation of
black hole accretion in Y15. Top panels: density, temperature, Mach number. Bottom panels: radial, meridional, and rotational velocities normalized by Keplerian velocity $v_k(r_0)$. Value of density is shown in code units. The density and temperature are time-averaged values, while all three velocity components are taken from one snapshot.}
\label{fig:GRMHDdata}
\end{figure*}

Y15 demonstrate that winds occupy the region of $0^\circ \la \theta \la 45^\circ$ and inflows are in the region of $45^\circ \la \theta \la 90^\circ$. Figure \ref{fig:GRMHDdata} shows the azimuthal distributions of some wind properties, such as  velocity and temperature, at $10^3r_g$ obtained in Y15.  Given the diversity of these quantities at  different $\theta$, we consider various boundary conditions at $r=10^3 r_g$ (Table \ref{tab:anaypara}). For example, the boundary conditions of the ``hot\_bhg\_sup'' model, $v_{r0}=3v_k$, $T_0=10^{10}$ K, and $v_{\phi 0} = 0.7v_{k0}$, correspond to winds at $\theta=20^\circ$. Note that such a set of physical quantities will result in a Mach number exceeds unity at the boundary, thereby a corresponding supersonic solution is obtained as we will discuss later. Most of the models for hot winds are assigned with a rotational velocity of $0.5v_{k0}$. We have confirmed that moderate deviation from this value ($0.4-0.6v_{k0}$) did not alter the results qualitatively.

\subsubsection{Thin accretion disks}\label{sec:thindisk}

Wind properties from thin disks are different from those from hot accretion flows. Since we still do not have a comprehensive theoretical model for the production of winds from thin disks, we incorporate both observational and numerical simulation results to determine the wind properties at the boundary.

Observationally, a sample of over 50 AGNs is collected from \textit{Suzaku} observations to study the properties of disk winds \citep{gofford_etal15}. These winds are detected in a spatial range of $10^2-10^4 r_s$ from the central  black hole. A power-law relation between the AGN bolometric luminosity and the wind velocity is found,
\begin{equation}
v_w = 2.5 \times 10^4\left(\frac{L_\mathrm{BH}}{10^{45}\; \mathrm{erg\;s^{-1}}}\right) \mathrm{km\;s^{-1}}.
\label{eq:velcold}
\end{equation}
{Assuming $L_\mathrm{BH}=10^{44}-10^{45}\;\mathrm{erg\;s^{-1}}$ ($0.01-0.1L_{\rm Edd}$ for black hole mass of $M_{\rm BH}=10^8M_\odot$) implies wind velocities in the range of $v_w=0.25v_k - 2.5v_k$ or $0.008c - 0.08c$, where $c$ is the speed of light. We adopt values increasing consecutively from $0.2v_k$ to $2v_k$ as our boundary conditions of different models. But given the diversity of various observations of quasar wind, values beyond this range are also tested.}

Due to the difficulty of obtaining  temperature and rotational velocity of winds from observations, we use numerical simulations to determine the values of these quantities. \citet{Proga2000} have performed hydrodynamical simulations on radiation line-driven winds from geometrically thin and optically thick accretion disks of luminous AGNs. Their results suggest that temperatures of wind at $10^3r_g$ range from $10^4-10^6$ K. Motivated by the above discussions, the boundary conditions of the fiducial model of thin disk winds are set to be $v_{r0}=v_k$, $v_{\phi0}=0.5v_k$, and $T_0=10^5$K. Various values of $T_0$ and $v_{r0}$ are adopted for  other models (Table \ref{tab:anaypara}). The adoption of a relatively large rotational velocity of $v_{\phi0}=0.5v_{k0}$ for thin disk winds comes from the detailed radiation line-driven wind simulations (Wang et. al. in preparation) at $10^3r_g$. Given that thin disk winds may be predominantly driven from the inner region of the disk, the angular velocity at $r_0=10^3r_g$ could be small, hence we explore a parameter space which $v_{\phi0}$ is allowed to vary from $0.01-0.5v_{k0}$.

\subsection{Solutions}

We look for transonic solutions following the approach presented in \S\ref{sec:eq} with boundary conditions given in \S\ref{sec:anaybc}. We begin with a general analysis of the topology to the solutions in \S\ref{sec:top}, then present the detailed solutions for hot accretion flows in \S\ref{sec:BHpot} and \S\ref{sec:stellarpot}, and for thin disks in \S\ref{sec:thindisk}. Special attention will be paid to the influence of including galaxy potential on the solutions.

\subsubsection{Solution topology} \label{sec:top}
\begin{figure}[ht]
\epsscale{1.15}
\plotone{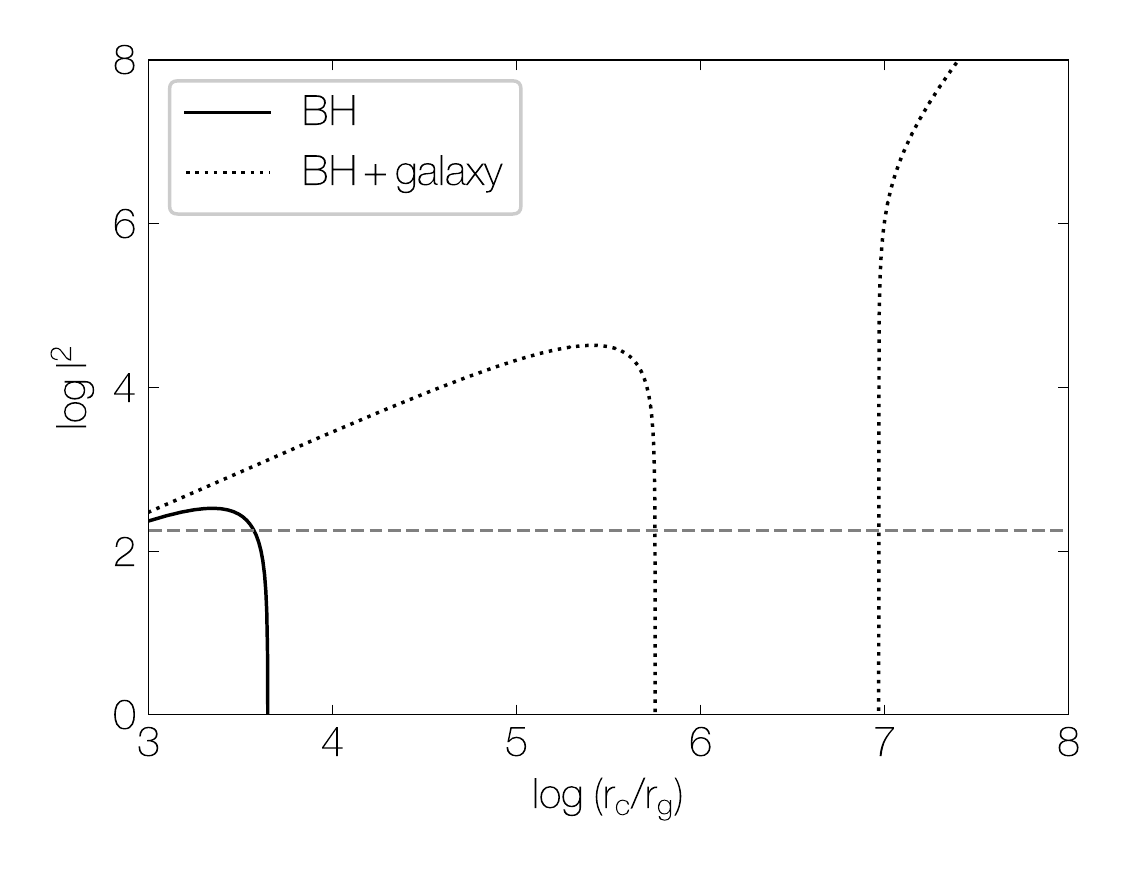}
\caption{The square of angular momentum $l^2$ as a function of the locus of the critical point $r_\textrm{c}$ at a specific Bernoulli parameter $E$. The solid curve corresponds to ``hot\_bh\_t'' (Table \ref{tab:anaypara}) which only includes the black hole potential, and the dotted curve corresponds to ``hot\_bhg\_tb'' which includes the potential of the black hole and the galaxy. The horizontal dashed curve shows the angular momentum $l^2$ adopted in ``hot\_bh\_t'' and ``hot\_bhg\_tb'' for easy inspection. The intersections with the dashed curve reveals the loci of critical points.}
\label{fig:l2}
\end{figure}

\begin{figure*}[ht]
\epsscale{1.15}
\plottwo{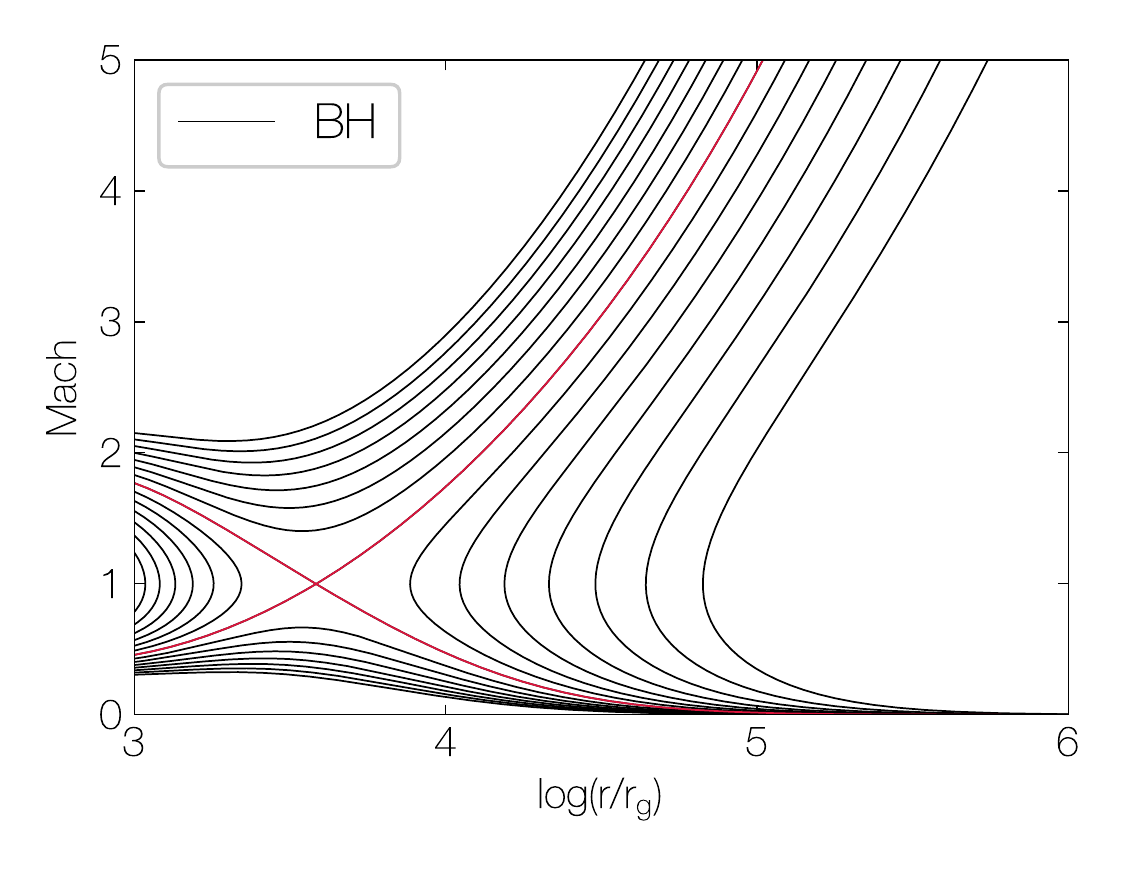}{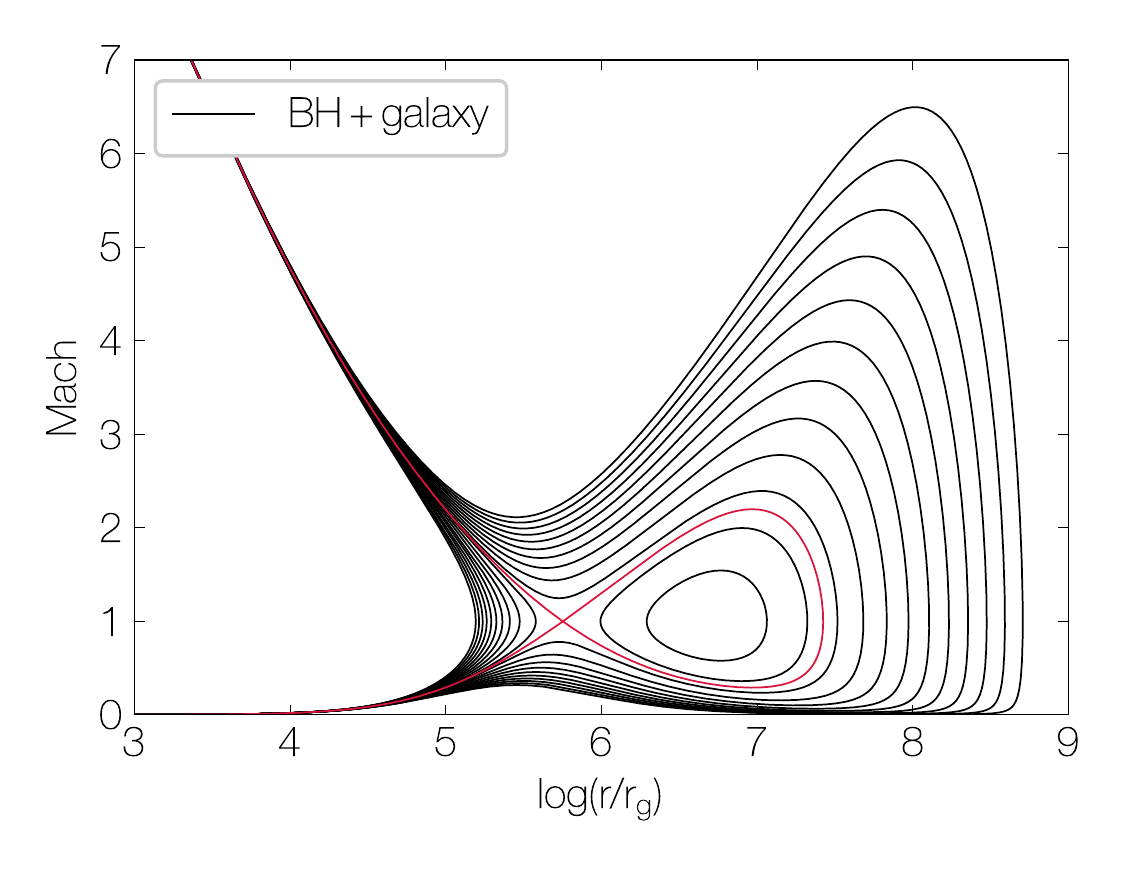}
\caption{Mach number of wind as a function of radius. Different curves in each panel correspond to the same mass flux $\dot M$ and angular momentum $l$ but different Bernoulli parameter $E$. The red curve in the left panel is for the ``hot\_bh\_t'' model, while the red curve in the right panel for the ``hot\_bhg\_tb'' model. Subsonic and supersonic solutions are present below and above the red curves.}
\label{fig:contour}
\end{figure*}

Based on Equation (\ref{eq:l2E}), Figure \ref{fig:l2} shows the square of angular momentum $l^2$ as a function of the locus of critical point $r_\textrm{c}$ at a specific Bernoulli parameter $E$. The solid and dotted curves correspond to model ``hot\_bh\_t'' and ``hot\_bhg\_tb'', respectively. On each curve, a specification of angular momentum $l$ marks the loci of the critical points, shown as intersections of the grey dashed curve or the solid and dotted curves.

Note that not all the critical points correspond to transonic solutions. Only saddle points are related to transonic solutions, while the vortex points only have mathematical meaning. These two types of critical points are usually denoted by ``X'' point and ``O'' point in the studies of hydrodynamical winds \citep{liang1980,lu1988}. The classic Bondi accretion has one critical point, which is a saddle point (\citealp*{bondi52}). When taking angular momentum into account, there are three critical points possible (\citealp*{abramowicz1981}). The point closest to the black hole corresponds to the accretion of material on to the black hole, which requires the gas to pass the sound speed eventually. Thus it is a saddle point and related to the transonic solution. The second point is a vortex point, and the third point is a saddle point which is relevant to the study of the wind here. In Figure \ref{fig:l2}, the intersection of the dashed curve and the black curve, and the first intersection of the dashed curve and the dotted curve correspond to the critical point of the wind. The first and second critical points of both the solid and dotted curves are located at radii $r<r_0$, thus not seen in the figure.

The second intersection between the dashed curve and the dotted curve is an additional point besides the aforementioned three critical points. Its presence is due to the inclusion of galaxy potential. The locus of this critical point is determined by the boundary condition. For example, we find the loci of the ``hot\_bhg\_ta'' model is much larger than that of the ``hot\_bhg\_tb'' model. This point is of vortex type as we will discuss in the following paragraph.

The topology of solutions is further illustrated in Figure \ref{fig:contour}, which shows the Mach number of wind as a function of radius. The red curve in the left panel is for the ``hot\_bh\_t'' model, which possesses the topology of a saddle point. It corresponds to the black curve in Figure \ref{fig:l2} with parameters listed in Table \ref{tab:anaypara}. The red curve in the right panel of Figure \ref{fig:contour} is for the ``hot\_bhg\_tb'' model. This model includes the galaxy potential hence possessing both a saddle point and a vortex point, which corresponds to the dotted curve in Figure \ref{fig:l2}. In both panels, the black curves are drawn by slightly adjusting the Bernoulli parameter $E$ to deviate from the transonic solution, but with $\dot M$ and $l$ fixed. The black curves above and below the red curve represent supersonic and subsonic solutions, respectively.

\subsubsection{Winds from hot accretion flow with black hole potential alone}\label{sec:BHpot}

\begin{figure*}[ht!]
\epsscale{1.0}
\plotone{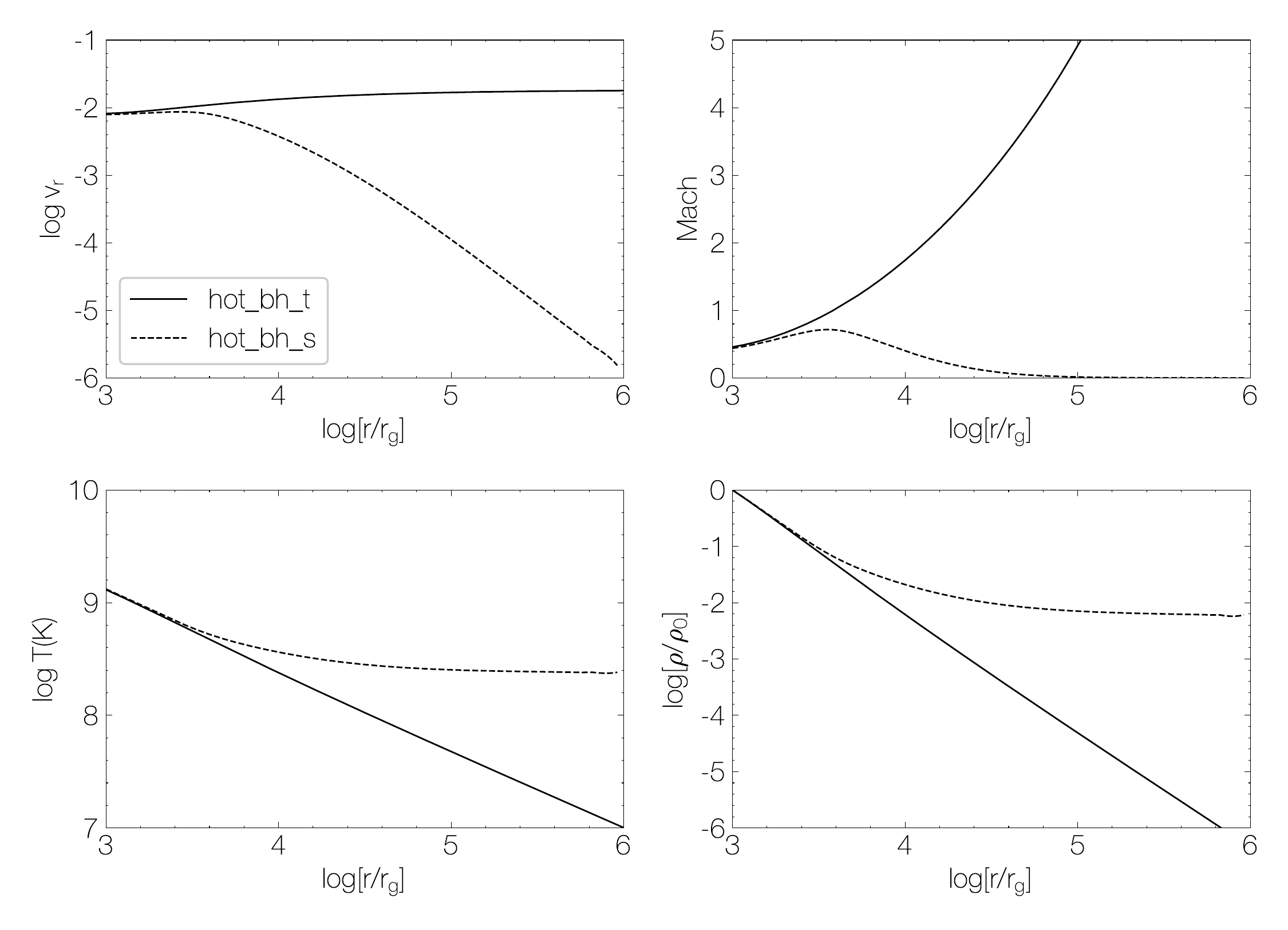}
\caption{Radial profiles of radial velocity, Mach number, temperature and density of the subsonic solution ``hot\_bh\_s'' (dashed) and transonic solution ``hot\_bh\_t'' (solid) of wind launched at $r_0= 10^3 r_{s}$, with detailed boundary parameters listed in Table \ref{tab:anaypara}.}
\label{fig:BHprofile}
\end{figure*}

\begin{figure}
\epsscale{2.2}
\plottwo{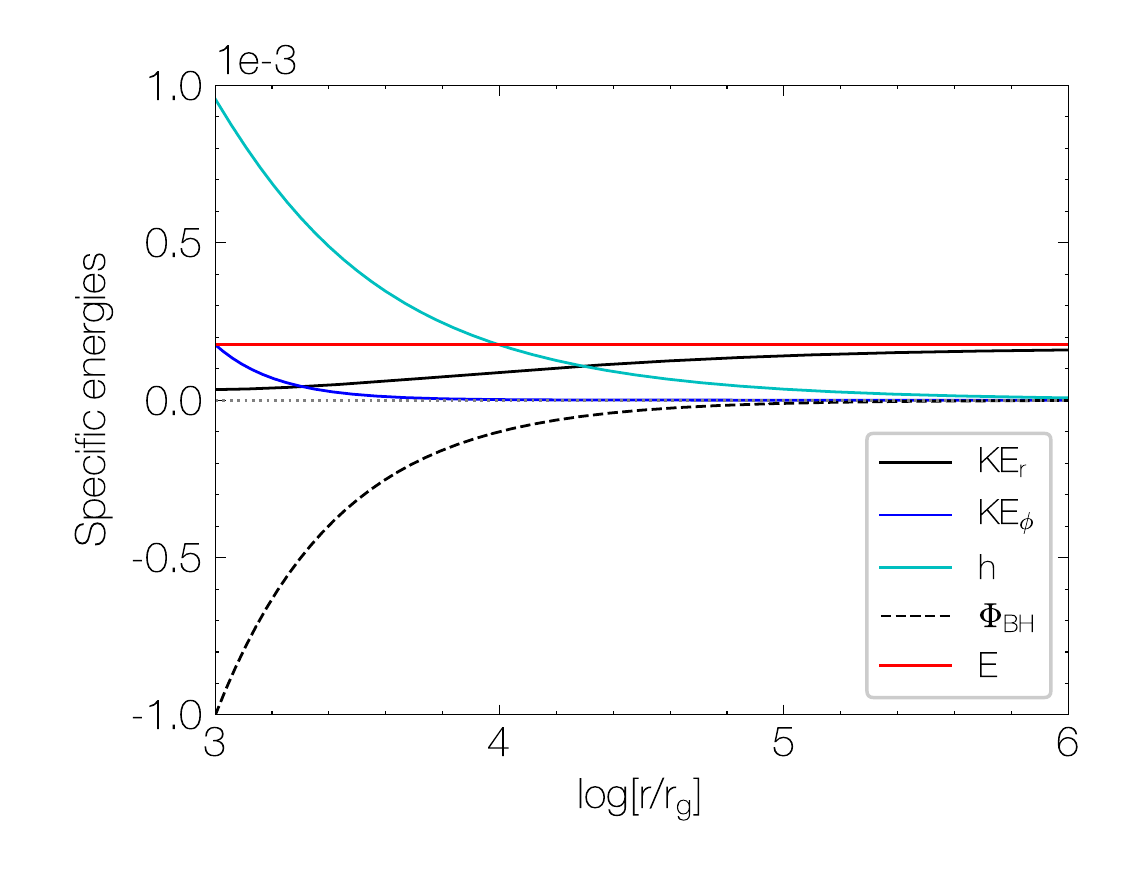}{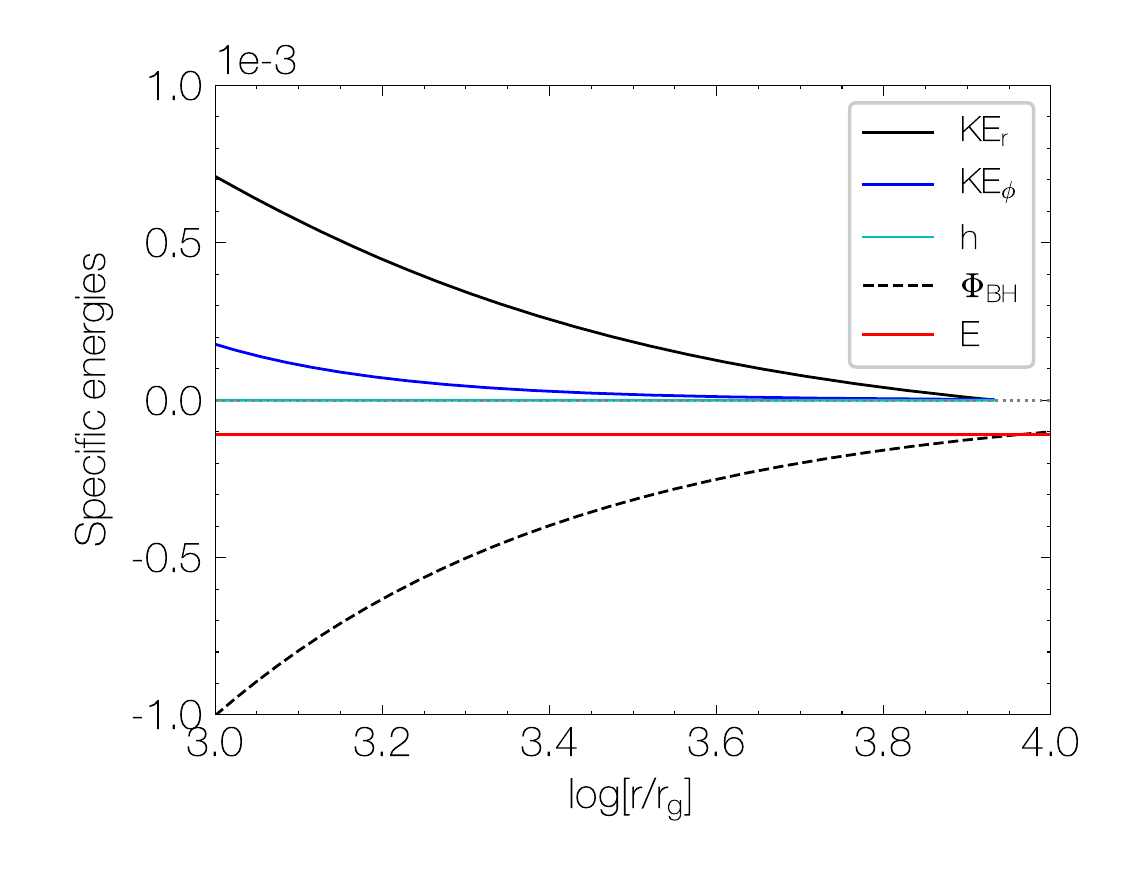}
\caption{Decomposition of the Bernoulli constant $E$ into individual components for model ``hot\_bh\_t'' (left) and ``cold\_fid' (right), including specific radial kinetic energy $\rm{KE}_r$ $(1/2v_r^2)$, specific rotational energy $\rm{KE}_{\phi}$ $(1/2v_{\phi}^2)$, specific enthalpy $h = c_s^2/(\gamma-1)$, and specific potential energy $\Phi_{\rm BH} = -GM/(r-r_s)$. For the right plot, a curve showing the $\sigma^2{\rm ln}r$ term in Equation 5 is also shown, which almost overlaps with the enthalpy curve. }
\label{fig:Ecomp}
\end{figure}

\begin{figure*}
\epsscale{1.0}
\plotone{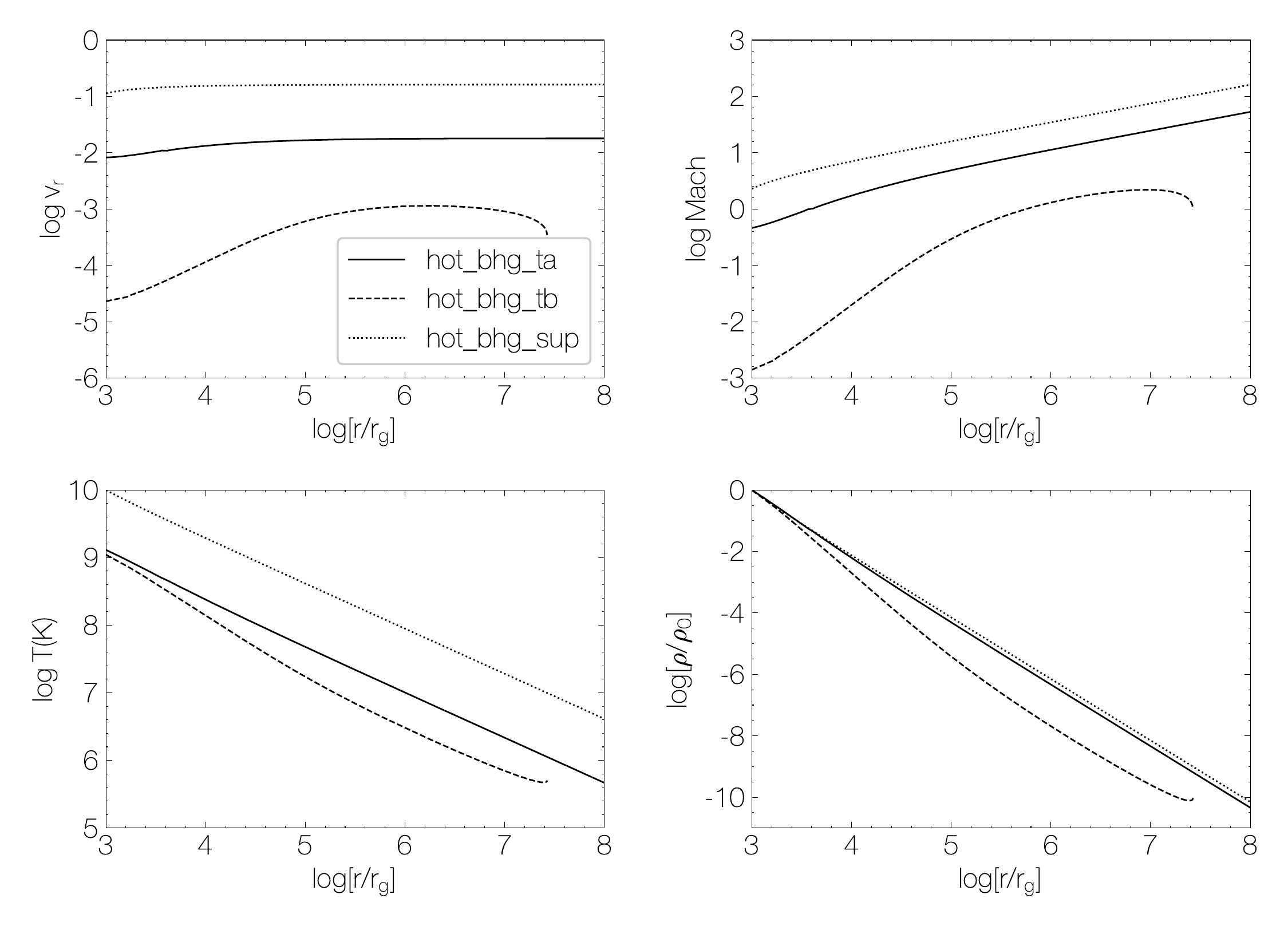}
\caption{Radial profiles of radial velocity, Mach number, temperature and density of the transonic solutions`` hot\_bhg\_ta'', ``hot\_bhg\_tb'' and the supersonic solution ``hot\_bh\_sup'', with detailed boundary parameters listed in Table \ref{tab:anaypara}.}
\label{fig:BHGprofile}
\end{figure*}

Figure \ref{fig:BHprofile} shows the radial profiles of velocity, density, Mach number and temperature of the representative transonic (solid line) and subsonic (dashed line) solutions. For other transonic and subsonic solutions the results are similar once their boundary conditions are in the realistic range. The supersonic solution is not shown here since its profiles share great similarity to the transonic solution (refer to the dotted curves in Figure \ref{fig:BHGprofile}).

It is clear from the transonic and supersonic solutions that the wind can escape from the black hole and propagate to very large radius. For the transonic solutions, we  find that the Mach number increases rapidly with radius, and the radial velocity only slightly increases. The radial profiles of density and temperature can be derived from the continuity equation and the polytropic relation. They can be roughly approximated by
\begin{align}
v_r \approx v_{r0}, ~~ \rho \approx \rho_0 (r/r_0)^{-2}, ~~ T \approx T_0 (r/r_0)^{-2(\gamma-1)}.
\label{transonic}
\end{align}

One may imagine that the radial velocity may decrease due to the gravitational force of the black hole. To understand why the radial velocity does not decrease with radius, we decompose the Bernoulli parameter $E$ into individual components (Equation \ref{eq:Be}) along the wind trajectory, the result is shown in the left plot of Figure \ref{fig:Ecomp}. It demonstrates that the increase of gravitational energy is mainly compensated  by the reduction of the specific enthalpy as well as the rotational energy.  At large radii, both the enthalpy and potential energy nearly vanish, and the total energy is dominated by the radial kinetic energy.  In other words, the gas pressure gradient and centrifugal forces overcome the gravitational force of the black hole and do work to accelerate the wind.

A constant or slightly increasing radial velocity resembles the velocity profile of wind on small accretion flow scales. On that scale, Y15 have found that the wind velocity almost keeps constant along their trajectories,  and it is also the specific enthalpy that compensates the increase of potential energy. However, the Bernoulli parameter of wind is found to increase along the wind trajectory rather then to keep constant. This is because the accretion flow is strongly turbulent, while conservation of Bernoulli parameter holds only for a strictly steady and inviscid flow. Our large-scale wind is well beyond the accretion disk scale so turbulence is reasonably assumed to be absent and the Bernoulli parameter should be constant.

The radial profiles of subsonic solution (``hot\_bh\_s'') strongly deviate from the transonic solution, as shown by the dashed curves in Figure \ref{fig:BHprofile}. The radial velocity rapidly decreases with radius, while the temperature and density drop slowly and tend to be constant at large distances. This subsonic solution is obtained by adjusting the Bernoulli parameter $E$ and keeping mass flux $\dot M$ and angular momentum $l$ the same as in ``hot\_bh\_t''. In this case, the sonic point condition Equation \eqref{eq:cs2} will not be satisfied. With a slight adjustment of $E$, the physical parameters at the boundary deviate from the parameters adopted in the transonic solution ``hot\_bh\_t''. As we have explained in \S\ref{sec:eq}, this will result in the failure of obtaining a transonic solution, implying that a transonic solution requires a specific combination of physical parameters at the boundary. 

The fine-tuning of the transonic solution may lead to the conclusion that the subsonic winds should be regarded as more physically realizable in nature rather than the supersonic winds. However, the evidence from the detection of solar winds objects this conjecture; solar winds detected are always supersonic at the Earth's orbit. The classic Parker wind model argues that the non-vanishing pressure in the subsonic solution prevents the wind from having a smooth transition to the interstellar medium at infinity, thereby excluding the subsonic solutions from physical solutions. Moreover, \citet*{velli94} states that the subsonic solutions are unstable to low-frequency acoustic perturbations, leaving transonic solution the only plausible solution to connect to infinity. Our 1D simulations prove the prevalence of transonic solutions by the facts that only transonic solutions are found, whereas no subsonic solutions are ever obtained (\S \ref{sec:icbc}).

\subsubsection{Winds from hot accretion flow with both black hole and galaxy potentials}\label{sec:stellarpot}

As the wind propagates toward large radii,  the galaxy potential should be included.  Figure \ref{fig:BHGprofile} shows the radial profiles of two transonic solutions ``hot\_bhg\_ta'' and ``hot\_bhg\_tb'', and a supersonic solution  ``hot\_bh\_sup''. The subsonic solution  ``hot\_bhg\_s'' is not shown in the figure since it resembles the profiles of the ``hot\_bh\_s'' model. As we have mentioned in \S\ref{hotboundary}, the wind at $r_0$ is a combination of components originating from various radii satisfying $r\la 10^3 r_g$, hence having various velocities.  The boundary conditions of the three solutions shown here mainly differ by their radial velocities, with the supersonic solution ``hot\_bhg\_sup'' having the largest velocity while ``hot\_bhg\_tb'' having the smallest one (Table \ref{tab:anaypara}). The velocity in the ``hot\_bhg\_tb'' model is extremely small and not realistic since we deliberately choose this value to manifest the effect of the galaxy potential. The radial velocity in ``hot\_bhg\_ta''  is close to the mass flux-weighted radial velocity of wind obtained in Y15.

The radial profiles of transonic solution ``hot\_bhg\_ta'' (solid line) are similar to those of the transonic solution of the black hole-potential-only case, i.e.,  the solid line in Figure \ref{fig:BHprofile}. The profiles of radial velocity, density, and temperature are also well approximated by Equation (\ref{transonic}). This is also the case of the supersonic solution ``hot\_bh\_sup''. This result indicates that the effect of galaxy potential is limited when employing the ``realistic'' boundary conditions.  We have estimated the largest distance the wind can propagate outward for ``hot\_bhg\_ta'' model and found that it is well beyond the scale of the galaxy. In reality, due to the interaction between wind and the interstellar medium, such an estimation should be regarded as an upper limit.

The effect of the galaxy potential will be significant if the radial velocity of wind is small at the boundary, as illustrated by the dashed curve in Figure \ref{fig:BHGprofile}, which  delineates the behavior of the solution ``hot\_bhg\_tb''. The wind has been accelerated first, with the radial velocity and the Mach number increasing. Beyond $r=10^6r_g$, the acceleration stops and the wind begins to decelerate beyond $10^7r_g$.  This is because the adopted  radial velocity at $r_0$ is so small that the galaxy potential plays a relatively much more important role. With the increase of radius, the galaxy potential energy becomes larger thus the kinetic energy has to be decreased due to the conservation of total energy. From this argument, we expect that the realistic transonic solution ``hot\_bhg\_ta'' shares the same behavior as ``hot\_bhg\_tb'' when our calculation domain is significantly extended, as the increasing galaxy potential will eventually slow the wind down.

The effect of the galaxy potential on wind dynamics depends on the magnitudes of the  gravitational force corresponding to this potential and also the velocity of wind itself. Small wind velocity and/or strong force will lead to prominent variations of wind velocity. In our case, though the force from the galaxy can be estimated, the amplitude of wind velocity requires detailed calculations with realistic boundary conditions. From the results presented above, we find that for most cases the wind velocity is not very small, and the effect of the galaxy potential is limited. 

From these results, we can obtain the following propagation scenario of wind launched from a hot accretion flow. Those winds close to the axis have the largest velocity at the boundary thus they will propagate to the largest distance in the galaxy, while those winds close to the equatorial plane have the smallest velocity thus they will stop at smaller distances.

\subsubsection{Winds from thin disks}\label{sec:thindisk}

\begin{figure*}[ht]
\epsscale{1.0}
\plotone{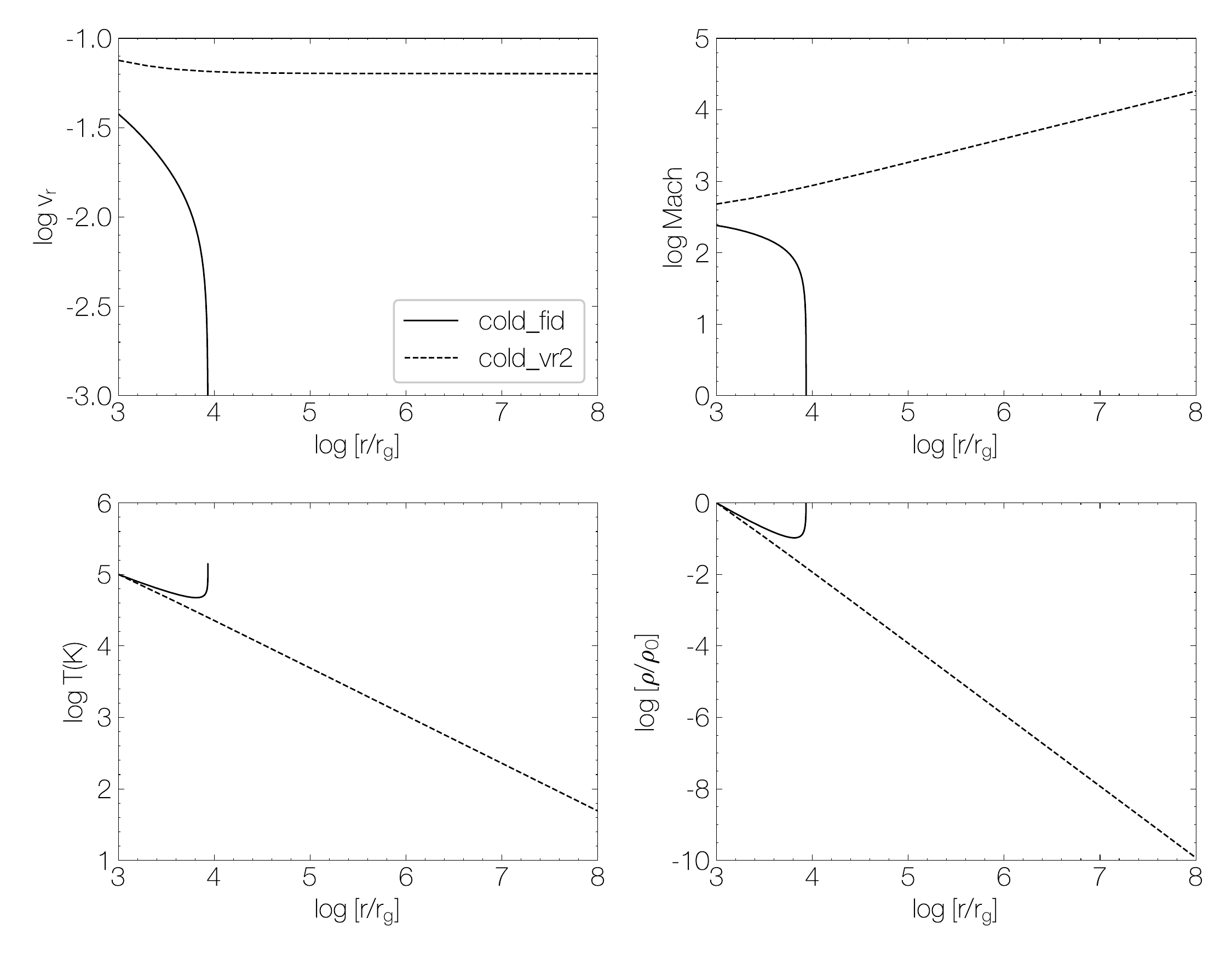}
\caption{Radial profiles of radial velocity, Mach number, temperature and density of thin disk winds for model ``cold\_fid'' and ``cold\_vr2''.}
\label{fig:coldpro}
\end{figure*}

From Table 1, we can see that the boundary conditions of winds from a thin disk are quite different from that of a hot accretion flow. The temperature and consequently the Bernoulli parameter are much lower in this case. In addition, for all the models of thin disk winds, the wind is already supersonic at $10^3 r_g$, mainly attributed to the low temperature of the wind. This implies that all the wind solutions can be realized in nature. 

Following the same approach as for the hot winds in studying wind dynamics, we find that in general the wind from a thin disk cannot propagate as far as in the case of hot accretion flows. The stop radii of different models depend on the value of the Bernoulli parameters, with a larger $E$ corresponding to a larger ``stop radius''. Figure \ref{fig:coldpro} shows the dynamics of two  example of models with different radial velocities taken from Table 1, ``cold\_fid'' and ``cold\_vr2''.  The difference of the velocity causes different Bernoulli parameters, with the Bernoulli parameter of the ``cold\_fid'' model being negative while the ``cold\_vr2'' model positive. We can see from the figure that the wind in the ``cold\_fid'' model even cannot escape from the black hole gravity and stops at $\la 10^4 r_g$. As we have emphasized before, since radiation and magnetic forces are neglected in our calculation, in reality the stop radius should be larger. The value of stop radius is sensitive to the value of Bernoulli parameter at the boundary. For the wind in the ``cold\_vr2'' model whose velocity at the boundary is only two times higher, it stops at a much larger radius, beyond the radial range shown in the figure. Given these results, the fact that winds from quasars have been widely observed as far as $\sim 15 $ kpc from the black hole  (e.g., \citealp{liu_etal13}) implies that either radiation and magnetic forces must continue to accelerate winds at large radii, or the velocity of wind at the boundary must be relatively large. This can potentially be checked by examining  observational data.

\begin{figure*}[ht!]
\epsscale{1.0}
\plotone{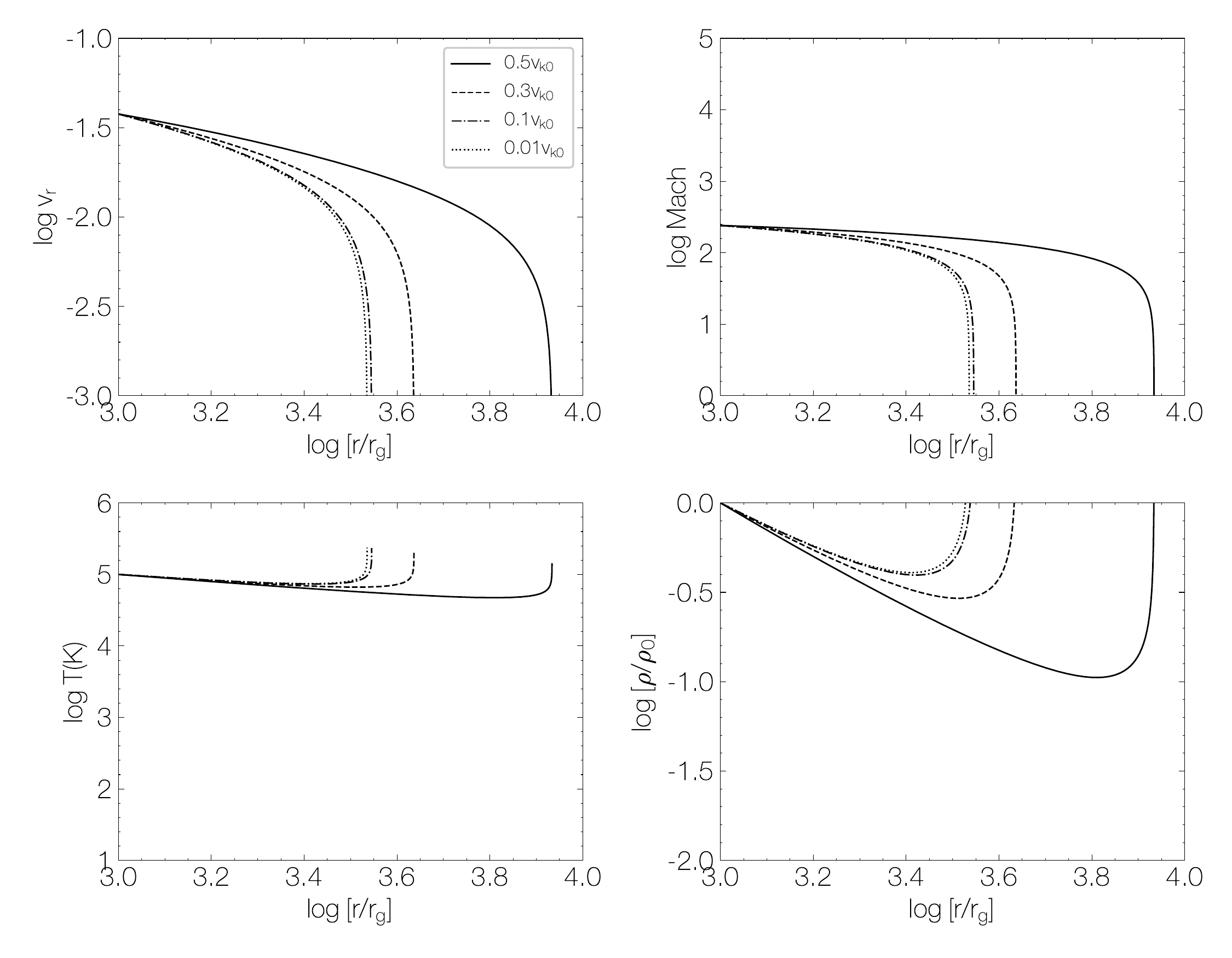}
\caption{Radial profiles of radial velocity, Mach number, temperature, and density profiles of different $v_{\phi0}$ for thin disk winds (Table \ref{tab:anaypara}). Curves denote angular velocities of $0.5v_{k0}$ (cold\_fid), $0.3v_{k0}$ (cold\_vphi0.3), $0.1v_{k0}$ (cold\_vphi0.1), and $0.01v_{k0}$ (cold\_vphi0.01) at the inner boundary. }
\label{fig:thinvphi}
\end{figure*}

 The value of stop radius can be roughly estimated as follows.  From Equation (5) we can deduce that  the radius where the wind stops is determined by equating the sum of two potential energy terms to the Bernoulli parameter, since the kinetic energy approaches zero there while the enthalpy is always negligible for the cold wind. As an illustrative example, we draw the right plot of Figure \ref{fig:Ecomp}, which shows the energy decomposition of the ``cold\_fid'' model.  At large radii, all the terms in Equation (5) are close to zero except the Bernoulli parameter and the black hole potential terms. These two non-zero terms are equal to each other at $\la 10^4 r_g$, which is exactly  the radius where the wind stops. We also vary the radial and rotational velocities as well as the temperature at the boundary for thin disk winds (Table 1). Slower radial and rotational velocity, and lower temperature result in a smaller stop radius, as the Bernoulli parameter is consequently smaller. In Figure \ref{fig:thinvphi}, it can be seen that when $v_{\phi0}$ is less than $0.1v_{k0}$, its impact on wind dynamics is weak as other parameters become dominant in determining the Bernoulli constant.
 
The above calculation results indicate that if the winds produced from the accretion disk have different components with different values of Bernoulli parameter, as likely the case, they will stop at different radii. This is similar to the case of wind from hot accretion flows, except that the stop radius should be systematically smaller due to the smaller enthalpy of cold wind. Those with a small stop radius, i.e., the so-called ``failed wind'', are invoked to explain the origin of broad line region of AGNs by \citet{proga19}.

Another question is how the velocity of winds changes with radius when they propagate outward. Such an information is useful for the AGN feedback study because numerical simulations of feedback having different spatial resolutions need to adopt the velocity of wind at different radius.  We can see from Figure \ref{fig:coldpro} that before the wind stops, its radial velocity only slightly decreases at the beginning, and then almost keeps constant throughout the radius. This result indicates that during the outward propagation of the wind, until close to the ``stop radius'', the rotational energy of wind almost exactly compensates the gravitational potential energy. This a result is very similar to those of a transonic solution ``hot\_bhg\_ta'' shown in Figure 6. Hence all the physical quantities can be  approximated by Equation \eqref{transonic}.

\section{Numerical Simulations}
Using the Athena++ code, we perform 1D and 2D numerical simulations to examine the time-dependent solutions of hydrodynamical winds and  compare with the analytical solutions. The conservation laws of mass,  momentum and energy in their conservative form read
\begin{gather}
\pdv{\rho}{t} + \nabla \cdot{(\rho \vb{v})} = 0, \\
\pdv{(\rho\vb{v})}{t}+ \nabla \cdot\left(\rho\vb{v}\vb{v}+P\vb{I}\right) = - \nabla \Phi, \\
\pdv{E}{t}+\nabla \cdot{\left[(E+P)\vb{v}\right]} = 0,
\end{gather}
where the total energy density is given by $E=P/({\gamma-1})+\rho\Phi+\rho v^2/2$ and $\vb{I}$ is the identity tensor.

\subsection{1D simulations}
We first discuss  the initial and boundary conditions and the polytropic index, two factors that deserve exceptional care before presenting the simulation results.

\subsubsection{Initial and boundary conditions} \label{sec:icbc}

For clarity, we start with 1D simulations without angular momentum ($v_\phi=0$). Thus, three variables are needed to be specified at the inner radial boundary, namely, pressure, density, and radial velocity. For the first attempt, $P, \rho, v_r$ are fixed in the ghost zone during the course of the simulation, which results in unphysical solutions. Discontinuities emerge in the first grid cell of the active zone, though it is connected with a smooth transonic solution at larger radii.

These solutions are interpreted as unphysical because the discontinuity clearly violates the conservation of mass. Mathematically speaking, this unphysical behavior arises because the number of dependent variables prescribed at the boundary exceeds the number of ingoing characteristics. Here by ingoing we mean that the characteristics point toward the computational domain (for details, see e.g. \citealp{wh84,gb00}). For 1D hydrodynamical problem, there exist three characteristics, relating to the entropy wave ($\lambda =v_r$), the sound waves propagating forward and backward ($\lambda = v_r+c_s$, $\lambda = v_r-c_s$). Here, $\lambda$ represents the eigenvalue and $c_s^2=\gamma P/\rho$ is square of the adiabatic sound speed. For the injection of subsonic winds at the inner radial boundary, the entropy wave and the forward sound wave have positive eigenvalues (speeds), while the backward sound wave always moves in the opposite direction to the wind. Thus, only two variables can be arbitrarily prescribed at the inner boundary, with the third one being adjusted itself.

We adopt time-dependent boundary conditions for which the pressure and density are fixed while the radial velocity are allowed to be adjusted at each time step. Specifically, only at the first time step, we set the values of three quantities at both the ghost zone and the first active grid the values given in Table \ref{tab:anaypara}. But at later times, only the pressure and the density are specified while the radial velocity is determined via enssuring mass conservation between the ghost gzone and the first active grid cell, i.e., $v_{r}^\prime = \rho_{0} v_{r0} r_{0}^2/\rho^\prime r^{\prime2}$, where subscript "0" denotes quantities in the first active grid cell and primed quantities are in the ghost cell.

\subsubsection{Polytropic index}\label{sec:gamma}

Mathematically, it can be proved that the set of equations comprised of conservation laws of mass and momentum, and the polytropic relation has transonic solutions when $\gamma < 5/3$. This criterion is obtained by requesting the critical point to be a saddle point, and is equivalent to having two eigenvalues of opposite signs for the differential equation. Moreover, outflowing gas always transits from subsonic to supersonic, thereby the slope of the radial velocity or the Mach number must be positive. This further confines the polytropic index to $\gamma< 3/2$ \citep{parker63}. Our 1D simulations confirm these conclusions.

Physically, the polytropic index $\gamma=5/3$ describes the adiabatic flows. A value of $\gamma<5/3$ may result from the intricate interplay among thermal conduction, heating and cooling. For instance, the rather flat radial profiles of the electron temperature in the near-Sun solar wind are usually attributed to the fact that the electron thermal conduction is rather efficient at temperatures exceeding $\sim 10^6$~K\footnote{In the context of black hole accretion with extremely low accretion rates, \citet{foucart_etal17} find that thermal conduction is however dynamically unimportant. It may be because that the magnetic field in their simulations is mainly toroidal while the temperature gradient is radial. Since conduction is expected to run along field lines, it reduces the conductive heat flux significantly. Another reason is related to the closure model they adopt, in which turbulence provides an effective collisionality to the plasma. This effect can strongly suppress the conduction. Both mechanisms do not hold in our case.}. As a result, empirical values of $\gamma \approx 1.1$ were frequently adopted (see e.g., \citealp{usmanov_etal00}, and references therein). On the other hand, $\gamma$ was empirically determined to be $\sim 1.46$ in interplanetary space (e.g., \citealp{tfa95}), indicating the existence of some weak but non-negligible heating (e.g., \citealp{marsch06}).

Note that the criterion of $\gamma<3/2$ holds only when omitting the angular momentum of the wind. Once angular momentum is taken into account, the $\gamma<3/2$ constraint is not necessarily valid. The inclusion of angular momentum complicates the set of equations, and there has no simple criterion on the polytropic index under this condition. We have tested a set of polytropic indices in no angular momentum cases and inspect if these $\gamma$ values can result in transonic solutions when including angular momentum. It is found that the angular momentum further confines the range of $\gamma$ that is possible to obtain the transonic solution.

\subsubsection{Simulation results}

In the simulation setup, we adopt the same boundary conditions as in analytic solutions. The exact way has been described in \S\ref{sec:icbc}. The wind is injected in the ghost zone cells, and the standard outflow boundary condition is employed at the outer boundary. We find that the transonic solution is the only steady solution no matter what initial values of radial velocity are specified. The subsonic solution obtained in analytical solutions are not present in numerical simulations. The explanation of its absence is based on the stability analysis \citep{velli94}, which demonstrates that the subsonic solution is unstable to low-frequency acoustic waves. The power-law fittings of transonic solutions obtained in numerical simulations with rotational velocity are fully consistent with the analytical solutions.

\subsection{2D simulations}\label{sec:2d}

Winds at different angles $\theta$ may interact with each other affecting their dynamics. This motivates us to perform 2D axisymmetric simulations to mimic more realistic conditions. The polytropic index $\gamma=4/3$ is adopted throughout, and the boundary conditions are employed from the small-scale 3D GRMHD simulation of black hole hot accretion flows (Y15) at a radius $r = 10^3r_g$ (Figure \ref{fig:GRMHDdata}). The physical quantities are averaged over azimuthal angle $\phi$. We only take the time-average for density and temperature over the time interval when the system reaches steady state. For velocities, we directly adopt values in one snapshot because the flow is turbulent in this simulations and the radial velocity frequently flips sign, so averaging over time will lead to significant underestimate due to the cancellation of positive and negative values. We emphasize that the data from $\theta\in[45^{\circ},90^{\circ}]$ are unphysical. The gas in that region is not wind but outflowing gas due to the outward angular momentum transport of the accretion flow. We include this region in 2D simulations but do not consider it when analyzing the simulation data.

\begin{figure}[ht!]
\epsscale{1.2}
\plotone{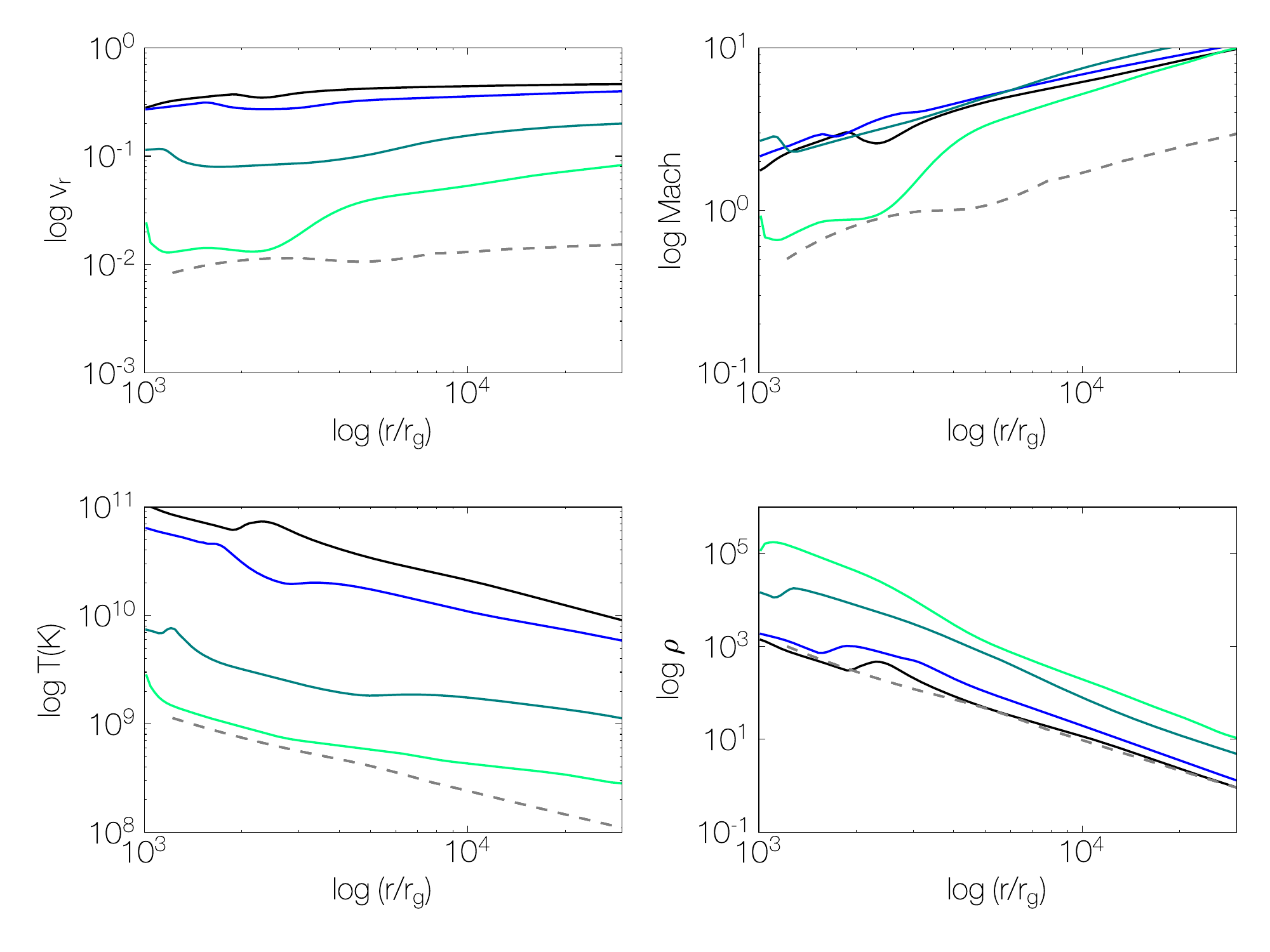}
\caption{Radial profiles of radial velocity, Mach number, temperature, and density profiles at several
different $\theta$ angles in the 2D simulation. Colors denote angles of $5^\circ$ (black), $10^\circ$ (blue), $20^\circ$ (dark green), and $30^\circ$ (light green). The analytic solution of model ``hot\_bhg\_ta''  is shown by dashed curves. }
\label{fig:2dsim}
\end{figure}

Figure \ref{fig:2dsim} shows the radial profiles of radial velocity, Mach number, temperature, and density of wind at different $\theta$. The wind reaches steady state up to $r\sim3\times10^4r_g$. The plot shows two types of behavior that are analogous to analytical solutions. First of which are flows at $\theta=30^\circ$ with subsonic radial velocity at the inner boundary. They are accelerated to become supersonic. This resembles the transonic analytical solution such as ``hot\_bh\_t'' or ``hot\_bhg\_ta''. For easy comparison, we show analytic solution of ``hot\_bhg\_ta'' by dashed grey curves. The profiles of density, temperature, and velocity are also well approximated by Equation \eqref{transonic}. Moving closer to the polar axis, supersonic solutions with radial velocity being supersonic at the boundary are shown, which correspond to ``hot\_bhg\_sup''. Its density, radial velocity, and temperature profiles also match well with the analytic solutions. The similarity between 1D and 2D simulations indicates that the gas motion at different $\theta$ has little interaction with each other, hence 1D simulation is a good approximation for wind dynamics. The results are also consistent with the conclusion of Y15 that the wind is largely laminar. We also find the wind trajectories are well approximated by straight lines, which justifies the assumption adopted in our analytical solutions.

\section{Summary and discussions}

Many works have been conducted to study the launching and acceleration of disk winds in the scale of black hole accretion flows. But it remains unclear how far the wind  can propagate, and what is the large-scale  dynamics of these winds. This is especially the case for winds from hot accretion flows, since the study of the hot wind is relatively new and winds are invoked to play an important role in AGN feedback. As the first paper of this series to answer these questions,  in this paper we  study the large-scale dynamics of wind under the most basic thermal assumption. The role of magnetic field will be investigated in our second paper.  In this work, both analytical study and numerical simulation are performed. The inner boundary is set to be $r_0=10^3r_g$. The boundary conditions of wind, which plays a crucial role in determining the wind dynamics, are discussed and presented in Table \ref{tab:anaypara} (for analytical models and 1D simulations) and Figure \ref{fig:GRMHDdata} (for 2D simulations). For hot accretion flows, the wind production in the accretion scale has been well studied by 3D GRMHD simulations, so these boundary conditions are taken from these simulations. For thin disks, theoretical studies which take into account all physical driving mechanisms are still lacked due to the technical difficulties, so we combine theoretical and observational studies to obtain the realistic boundary conditions. Specifically, the velocity of wind is taken from observations while the temperature and rotational velocity are from numerical simulations. The potential energy of the black hole as well as the host galaxy are taken into account. We focus on transonic and supersonic solutions as they are physical solutions that can be realized in nature. Mathematically, subsonic solutions also exist, but the  solution is  unphysical, partly because the subsonic solution is unstable.

For winds from hot accretion flows, we find that winds can propagate to very large radius. Transonic, supersonic, and subsonic analytical solutions have all been obtained, depending on boundary conditions specified. The physically viable transonic and supersonic solutions show similar radial profiles of dynamical quantities (Figure \ref{fig:BHprofile} and Figure \ref{fig:BHGprofile}), and they are approximated by Equation \eqref{transonic}. We find that when the galaxy potential is included the solution does not deviate much from models excluding it. The gravitational force exerted by the host galaxy is not able to modify the wind poloidal velocity significantly. This is mainly because the wind velocity from our detailed calculation is very large compared to the acceleration by the galaxy. The radial velocity of wind keeps constant with increasing radius, because the enthalpy and rotational energy of winds almost exactly compensate the potential energy when winds propagate outward (Figure \ref{fig:Ecomp}).

For wind from thin disks, we find that the wind usually stops at a smaller radius compared to the wind from hot accretion flows. This is mainly because the temperature of wind is much lower thus the Bernoulli parameter is smaller. The stop radius is determined by equating the Bernoulli parameter to the potential energy, since all other terms in the Bernoulli equation can be neglected at the stop radius. Before the stop radius, however, same with the hot wind the radial velocity of cold wind also roughly keeps constant and the radial profiles of physical quantities can also be roughly described by Equation \eqref{transonic}. During the propagation of cold wind, different from the hot wind, only the rotation energy compensates the increase of gravitational potential energy since the enthalpy is negligibly small. 

We have performed 1D and 2D numerical simulations to study the dynamics of hot wind and  compared with analytical solutions. Starting with a subsonic solution results in one of three characteristics having a negative velocity leading to wave propagation out of the simulation domain. Hence, only two variables among $P,\rho$ and $v_r$ can be given at the inner radial boundary, and the other one should be self-consistently determined  to satisfy the compatibility requirements. This is realized through time-dependent boundary condition where radial velocity is allowed to be adjusted at each time step via ensuring mass conservation. We only find the transonic and supersonic solutions; the subsonic solution found in the analytical solution is not present because the solution is unstable. The profiles of dynamical quantities for the transonic and supersonic solutions are well consistent with those from analytical solutions.

In this study, we have not taken into account the interaction between winds and interstellar medium (ISM). An interesting question is then how far the wind can propagate when ISM is included. To estimate the largest distance, we equate the ram pressure of wind to the thermal pressure of ISM. The thermal pressure of wind can be ignored since at large radii it is supersonic. We adopt the observed values of mass flux and velocity of thin disk winds \citep{gofford_etal15}. The mass flux is 2.5 $\rm M_\odot \:yr^{-1}$ and the wind velocity is $v_r=1.5v_{\rm K}$ at $r=10^3 r_g$, for black hole mass $\rm M_{BH}=10^9M_\odot$ and $L_{\rm BH}/L_{\rm Edd}=0.1$. The solid angle of wind is assumed to occupy $\sim 30\%$ of the whole sphere \citep{tombesi_etal11}. The terminal distance of wind is estimated to be
\begin{equation}
D_{\rm term} \approx 65 \times\left( \frac{10^{-3} \; \mathrm{cm^{-3}}}{n_\mathrm{ISM}}  \right)^{1/2} \left( \frac{10^{7} \; \mathrm{K}}{T_{\rm ISM}}  \right)^{1/2}  \; \rm kpc,
\label{eq:term}
\end{equation}
where $n_{\rm ISM}$ and $T_{\rm ISM}$ are the number density and temperature of the ISM, respectively. Taking $n_{\rm ISM} = 10^{-3}\;\rm cm^{-3}$ and $T_{\rm ISM}=10^7\;\rm K$ results in a distance of about $65$ kpc, consistent with observations of quasar-driven winds within a factor of a few (e.g., \citealp{liu_etal13}).  

We emphasize that Equation \eqref{eq:term} should be regarded as the upper limit of the terminal distance. When winds propagate outward, they will be contaminated by the gas in the ISM, leading to the decrease of radial velocity due to conservation of  momentum, and the terminal distance will become smaller. The magnitude of  decrease depends on the contrast between momentum flux of wind and the mass of ISM gas picked up in winds.

\section{Acknowledgements}

This work is supported in part by the National Key Research and Development Program of China (Grant No. 2016YFA0400704), the Natural Science Foundation of China (grants 11573051, 11633006, 11650110427, 11661161012), the Key Research Program of Frontier Sciences of CAS (No. QYZDJSSW-SYS008), and the Astronomical Big Data Joint Research Center co-founded by the National Astronomical Observatories, Chinese Academy of Sciences and the Alibaba Cloud. This work made use of the High Performance Computing Resource in the Core Facility for Advanced Research Computing at Shanghai Astronomical Observatory.

\end{document}